\documentclass[onecolumn,superscriptaddress]{revtex4}

\usepackage{ifpdf}
\usepackage{graphicx}
\usepackage[usenames,dvipsnames]{color}

\DeclareGraphicsExtensions{.pdf,.png,.eps,.jpg}

\begin{document}

\title{Combined large spin-splitting and one dimensional confinement in surface alloys}

\author{Alberto Crepaldi}
\address{Institut de Physique
de la Mati\`ere Condens\'ee (ICPM), Ecole Polytechnique
F\'ed\'erale de Lausanne (EPFL), Station 3, CH-1005 Lausanne,
Switzerland}

\author{Gustav Bihlmayer}
\address{Peter Gr\"unberg Institut and Institute for Advanced Simulation, 
Forschungszentrum J\"ulich and JARA, D-52425 J\"ulich, Germany} 

\author{Klaus Kern}\address{Institut de Physique de la
Mati\`ere Condens\'ee (ICPM), Ecole Polytechnique F\'ed\'erale de
Lausanne (EPFL), Station 3, CH-1005 Lausanne, Switzerland}
\address{Max-Plank-Institut f\"ur Festkörperforschung,
D-70569, Stuttgart, Germany}

\author{Marco Grioni}
\address{Institut de Physique de la Mati\`ere Condens\'ee
(ICPM), Ecole Polytechnique F\'ed\'erale de Lausanne (EPFL),
Station 3, CH-1005 Lausanne, Switzerland}
\date{\today}


\begin{abstract}
We have found and characterized by angle-resolved photoelectron spectroscopy (ARPES) quasi-one dimensional spin-split states in chain-like surface alloys formed by large \emph{Z} elements (Bi and Pb) at the Cu(110) surface. The ARPES results are supported by first-principles relativistic calculations, which also confirm the spin polarization of these states, characteristic of the Rashba-Bychkov effect.
The Fermi surface contours are open, but warped, as a result of the interaction with the bulk Cu conduction band.
This interaction introduces a $k$ dependence of the spin splitting perpendicular to the chains direction.
We have also investigated the influence of the atomic spin-orbit parameter in substitutional isostructural $\mathrm{Bi_{1-x}Pb_{x}}$ overlayers, and found that the magnitude of the spin splitting can be continuously tuned as a function of stoichiometry.
\end{abstract}

\maketitle


\section{Introduction}

Inversion symmetry breaking in the presence of strong spin-orbit coupling (SOC) leads to the lifting of the spin degeneracy of the electronic states. The theory of this effect in non-centrosymmetric bulk crystals was developed by Dresselhaus \cite{Dresselhaus_1955} and Rashba \cite{Rashba_1960}. The predicted spin-splitting was later experimentally verified in semiconducting quantum wells by photocurrent \cite{Ganichev_PRL_2004} and magnetoresistance measurements \cite{Hassenkam_RPB_1997}. Only recently it became possible to directly observe it by angle-resolved photoelectron spectroscopy (ARPES) in the bulk semiconductor BiTeI \cite{Ishi_NatMat_2011, Crepaldi_PRL_2012}. The Rashba-Bychkov (RB) model describes a related effect, namely the spin degeneracy lifting at surfaces and interfaces, where inversion symmetry is naturally broken \cite{RB_1984}. Spin-split states were directly observed by ARPES first at the Au(111) surface \cite{Lashell_PRL_1996}, and later at other metal surfaces such as W(110) \cite{rot_PRL_1999, Miyamoto_PRL_2012}, Bi \cite{Koroteev_PRL_2004} and Ir(111) \cite{Vary_PRL_2012}. Spin-resolved ARPES has confirmed the predicted characteristic vortical spin texture, with the spin polarization (mostly) in-plane and perpendicular to the electron's wave vector {\em k} \cite{Rote_PRL_2002, hoesh_PRB_2004}. 

In the original RB model, the splitting of the spin-polarized bands is determined by the gradient of the surface potential, i.e. by the strength of the surface electric field along the surface normal. The model successfully describes the topology of the bands and of the Fermi surface, but 
grossly underestimates the magnitude of the effect. The quantitative discrepancy is dramatic for the $giant$~spin splitting discovered in the BiAg$_2$ surface alloy \cite{Ast_PRL_2007}, and in similar Bi- and Pb-based ordered interfaces formed at the (111) surfaces of noble metals (Cu and Ag) \cite{moreschini_PRB_2009, Bentmann_EPL_2009}) and semiconductors (Si and Ge \cite{Emm_PRL_2008, Gierz_PRL_2009, Sakamoto_PRL_2009, Sakamoto_PRL2_2009, Yaji_NatCom_2010}). Clearly the simple model misses some important aspects of the problem. More elaborate models
consider atomic contributions \cite{Cercellier_PRB_2006}, the in-plane anisotropy of the surface potential \cite{Premper_PRB_2007}, the asymmetry of the wavefunctions \cite{Bihlmayer_PRB_2007}, or again the existence of a polarization of the local orbital angular momentum \cite{Park_PRL_2011}. Quite generally, it is found that buckling of the
surface layer containing the heavy atoms ($\Delta z$) is conducive to larger spin splittings. This was recently confirmed by combined ARPES and IV-LEED experiments \cite{Gierz_PRB_2010}.
 
Previous reports of giant spin splittings were focused on fcc (111) surfaces, while no giant Rashba effect was observed at the corresponding, less symmetric, (110) surfaces. Theory predicts a large and anisotropic spin splitting e.g. for Au(110)  \cite{Nagano_JPCM_2009, Simon_PRB_2010}, but its surface state lies above $E_{F}$. Attempts to manipulate its binding energy cause also the broadening of the band, whose splitting becomes experimentally unresolved \cite{Reinert_PRB_2008}. Similar difficulties have been reported for the unoccupied spin split surface state of Pt(111) \cite{Frantzeskakis_PRB_2011, Bendounan_PRB_2011}.
 
In this article we present ARPES data and first-principles calculations for the Bi/Cu(110) $p(4\times1)$ interface. Previous structural studies indicate the formation of a substitutional alloy, with large buckling of the heavy atoms in substitutional sites \cite{Lottermoser_SS_1996, Nagl_PRB_1995}. We show that the anisotropy of the surface alloy yields spin polarized electronic states with quasi-one-dimensional (1D) character. This represents a novel discovery in the field of Rashba-Bychkov systems, and it extends previous reports of spin splitting in 1D Au chains grown on stepped Si surfaces \cite{Barke_PRL_2006, Sanchez_PRL_20004, Okuda_PRB_2010}, 1D Pt self-assembed wires on Si(110) \cite{Yeom_PRL_2013} and the spin polarized 1D state at the surface of the stepped Bi(114) \cite{Wells_PRL_2009}. We also explored the role of the atomic SOC in mixed $\mathrm{Bi_{1-x}Pb_{x}}$/Cu(110) surface alloys, and found that the magnitude of the spin splitting can be continuously tuned as a function of stoichiometry. 

In all cases, we find that the measured splitting of the spin-polarized bands is different for nominally equivalent wave vectors in different surface Brillouin zones (BZs). This a consequence of the interaction of the one-dimensional surface states with the substrate bulk bands. It indicates that the form of the surface state wave function determines the magnitude of the effective SOC, as previously proposed for BiAg$_2$ \cite{Bihlmayer_PRB_2007}. 


\begin{figure}[b!]
  \centering
 \includegraphics[width = 0.95\textwidth]{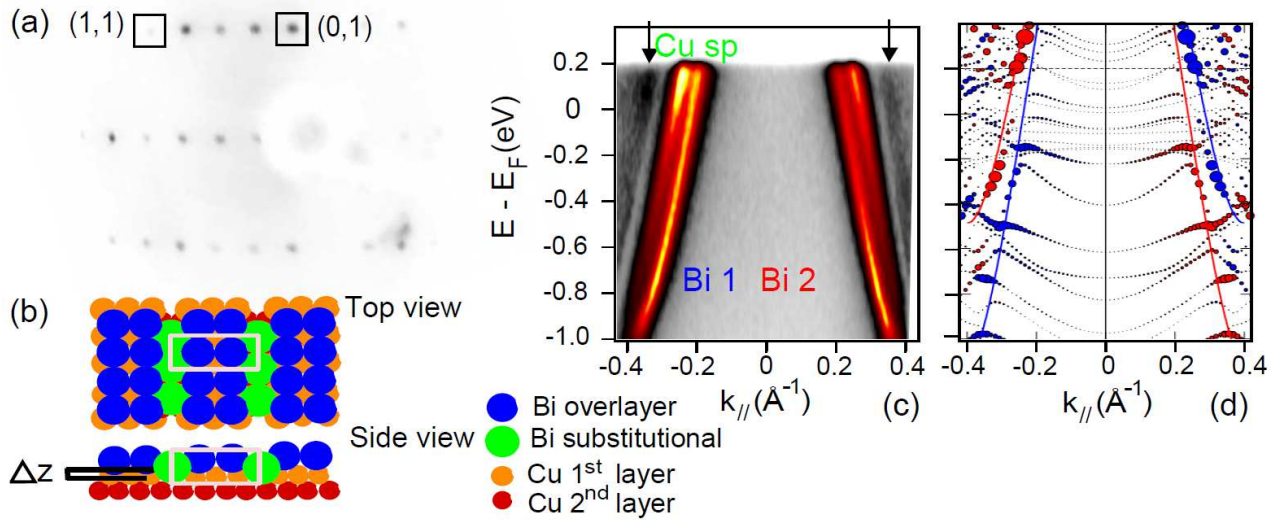}
  \caption[didascalia]{(a) LEED
  image of the Bi/Cu(110) $p(4\times1)$ interface. Black squares enclose the (0,1) and (1,1) substrate spots; (b) schematics of the surface structure from Ref. \cite{Lottermoser_SS_1996}. The white rectangle is the surface unit cell. The Bi atoms in substitutional sites form parallel chains, characterized by a large buckling ($\Delta$\emph{z} = 0.7~\AA) with respect to the neighboring Cu atoms.
  (c) Measured and and (d) calculated band dispersion parallel to the chains $\overline{\Gamma Y}$ direction.  Blue and red markers indicate opposite values of the spin polarization, perpendicular to the wave vector, and the size of the markers is proportional to the weight of the corresponding state at the interface between the film and the vacuum. A pair of spin split surface states (labelled Bi1 and Bi2) disperse with negative effective mass.
  }
  \label{fig:structure_arpes}
\end{figure}


\section{Experiment}

Cu(110) was cleaned by repeated sputtering (with Ar$^+$ at 300 K, 1keV for 30 minutes) and annealing cycles (900 K for 25 minutes). The quality of the surface was verified by low energy electron diffraction (LEED), which yielded sharp (1$\times$1) spots. Bi and Pb were evaporated (co-evaporated in the case of the mixed alloy) on the substrate kept at room temperature from a commercial EFM3 Omicron evaporator. For both kinds of ad-atoms, a 0.5 ML coverage yields a $p(2\times2)$ reconstruction. In the case of Bi, for higher coverage ($ \sim ~ 0.75~ML$) the $p(4\times1)$ reconstruction is formed, in agreement with the literature \cite{Lottermoser_SS_1996}. The phase diagram for the Pb case is more complicate, and several phases with various $p(n\times1)$ periodicities were observed, again in agreement with the literature \cite{Nagl_PRB_1995}. We consider here the higher coverage $p(5\times1)$ phase, at $ \sim ~ 0.8~ML$. The relative Bi and Pb content in the mixed alloy was determined from the intensity ratio of the $5d$ core levels measured by photoemission.  The structural quality of the surface alloy improved after a mild post-annealing at 500 K. ARPES measurements were performed with a hemispherical SPECS 150 Phoibos analyzer and a high brightness Gammadata VUV 5000 lamp operating at the HeI$\alpha$ ($21.2$ eV) (HeII$\beta$ at $48.1$ eV for the core levels). All measurements were performed at liquid nitrogen temperature, with the angular and energy resolution set respectively to 0.2$^{\circ}$ and 10 meV.\\

\section{Electronic structure calculation}
The calculations have been performed within density functional theory (DFT) in the
generalized gradient approximation (GGA)~\cite{PW91}. We used the full-potential linearized
augmented planewave method implemented in the {\sc Fleur} code~\cite{FLEUR}. The
Bi/Cu(110) $p(1 \times 4)$ surface was simulated by a symmetric film consisting
of 11 (for the bandstructures: 23) Cu(110) layers covered with Bi according to the
model of Lottermoser et al.~\cite{Lottermoser_SS_1996}. The planewave cutoff ($k_{\rm max}$)
and muffin-tin radii ($R_{\rm MT}$) are chosen to yield $R_{\rm MT} k_{\rm max} = 8.5$.
The atomic positions were relaxed until all forces were smaller than 26~meV/\AA. The
resulting structure compares well with the experimental data, e.g. the buckling within
the Bi layer and the lateral shifts are reproduced within 1-2 \%. Only the Cu-Bi
distances are overestimated, as GGA tends to overestimate the atomic volume of Bi.
Spin-orbit coupling was taken into account self-consistently~\cite{li_1990}, having only a minor
influence on the structure, while the electronic structure is substantially affected,
in particular states near the surface.


\section{Electronic properties of Bi/Cu(110) $\mathbf{p(4\times1)}$}

Figure~\ref{fig:structure_arpes}~(a) shows a LEED image of the Bi/Cu(110) $p(4\times1)$ interface. Black squares identify the (0,1) and (1,1) diffraction spots of the substrate. Figure~\ref{fig:structure_arpes}~(b) displays a schematic model of the Bi/Cu(110) interface, determined by surface x-ray diffraction \cite{Lottermoser_SS_1996}. The surface unit cell (white rectangle) contains two inequivalent Bi atoms. The first, at the corner of the cell, occupies a substitutional site in the topmost Cu layer, as shown in the bottom panel of Figure~\ref{fig:structure_arpes}~(b). By analogy with the Bi/Ag(111)~$(\sqrt3\times\sqrt3)$R30$^{\circ}$ system, this is the atom involved in the formation of the surface alloy \cite{Ast_PRL_2007}. 
The alloyed Bi atoms form parallel rows along the [001] direction, as also reported by an STM study of the related Pb/Cu(110) $p(n\times1)$ interfaces \cite{Nagl_PRB_1995}. Hereafter we refer to them as Bi chains. The spacing between adjacent chains is $1.02$~nm. Interestingly, the chains are buckled, with the Bi atom higher by $\Delta$\emph{z} = 0.7~\AA~~with respect to the neighboring Cu atoms \cite{Lottermoser_SS_1996}. This value is comparable the buckling of the (Bi, Pb) atoms in the surface alloys formed at the (111) surfaces of Ag and Cu \cite{Gierz_PRB_2010}. 

Figure~\ref{fig:structure_arpes}~(c) is an ARPES image of the band dispersion parallel to the Bi chains, along the high-symmetry $\overline{\Gamma\textmd{Y}}$ direction of the $p(4\times1)$ BZ ($\overline{\Gamma\textmd{Y}}$ = 0.86 \AA$^{-1}$). It shows three dispersive features, symmetric around the $\overline{\Gamma}$ point. The outermost band (indicated by black arrows) is the Cu bulk \emph{sp} conduction band. Two inner, more intense features, labelled Bi1 and Bi2, disperse with negative effective mass. They are suggestive of a pair of spin-split surface states, with a crossing point at $\overline{\Gamma}$ above the Fermi level. Similar split-states have been observed for the BiAg(110) $p(4\times1)$ interface \cite{Carbone_2010}. The spin splitting, obtained from the fit to the momentum distribution curve (MDC) at the Fermi level, is $\Delta$\emph{k} = 2$k_{0}$ = 0.048 $\pm$ 0.005 \AA$^{-1}$, where $k_{0}$ is the momentum offset of the band maximum. A parabolic fit of the two subbands yields an effective mass m$^{*} = -0.26 \pm 0.05$ m$_{e}$. 

Figure~\ref{fig:structure_arpes}~(d) illustrates the results of the DFT calculation. They reproduce very well the experimental dispersion of the Bi induced states, and confirm the characteristic Rashba-like spin polarization of the Bi states. The Cu(110) bulk continuum appears in the slab calculation as a manifold of discrete states which disperse with a positive mass and hybridize with the Bi bands.
Since the Bi states form a resonance with the Cu bulk bands, in a film calculation
they hybridize differently with different Cu quantum well states (QWS). This induces also a
spin-polarization of the Cu QWS with large weight at the film-boundary. The red
and blue lines in Fig. 1(d) are 4th-order fits to the resonances, revealing a
splitting $\Delta k = 0.044$~\AA$^{-1}$ at the Fermi level. A quadratic fit to the
Bi states in the range from 0.5~eV binding energy to the Fermi level leads to
an effective mass of -0.27 m$_{\rm e}$. The calculated electronic properties thus well fit to the experimental findings.
The calculated Cu states are rather flat in proximity of the $\Gamma$ point, while they display larger dispersion at larger $k$-vectors, thus well reproducing the experimentally observed dispersion of the Cu \emph{sp} state.

\begin{figure}[!t]
  \centering
  \includegraphics[width = 1.0\textwidth]{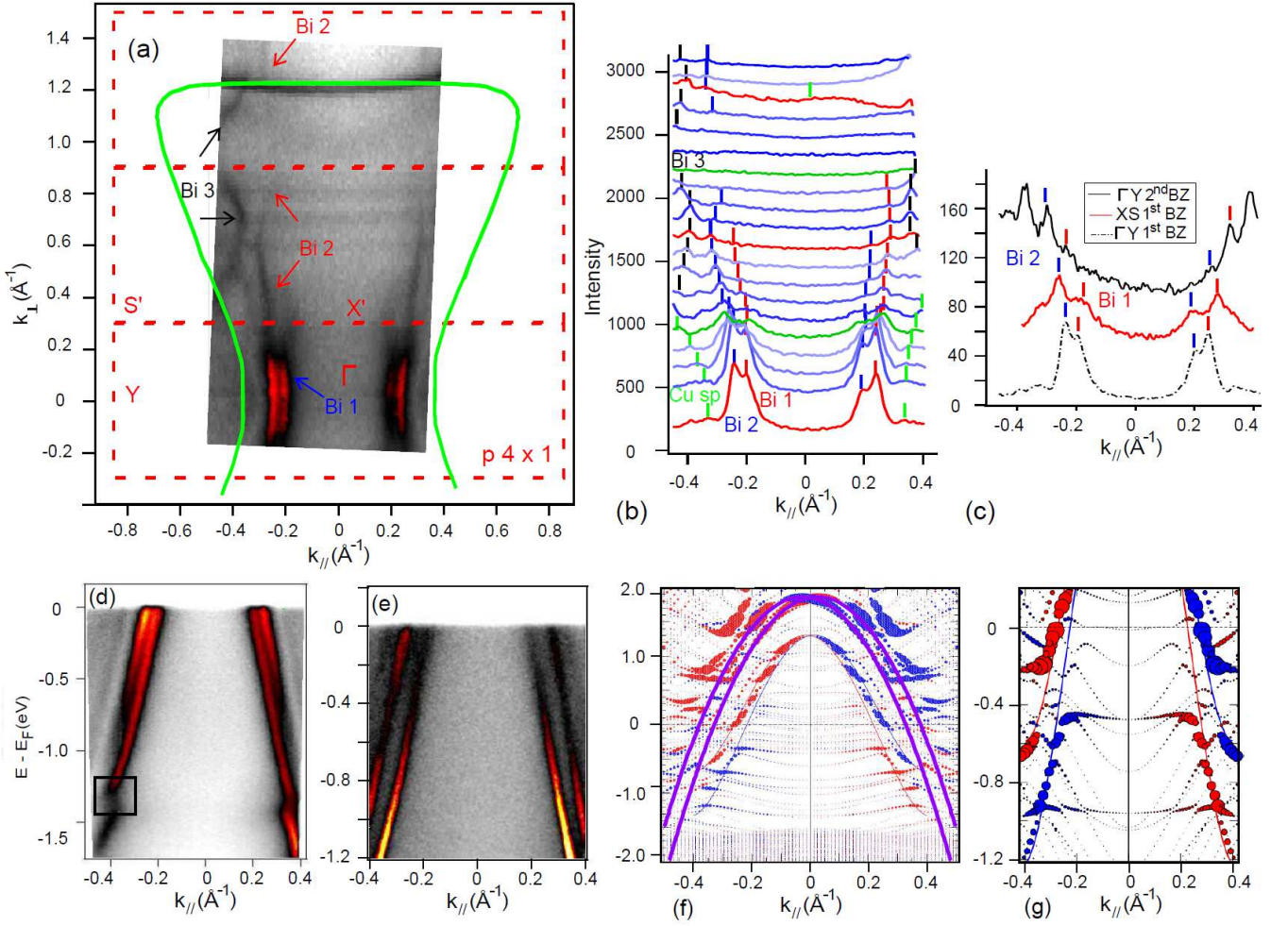}
  \caption[didascalia]{(a) ARPES constant energy contour for $E=E_F$. Red dashed lines delimit the $p(4\times1)$ BZ ($\overline{\Gamma\textmd{Y}}$ = 0.86 \AA$^{-1}$; $\overline{\Gamma\textmd{X}}$ = 0.30 \AA$^{-1}$). Blue and red arrows indicate the open contours of the quasi-1D Bi1 and Bi2 spin-split states. The green line outlines the FS contour of the bulk Cu $sp$~band.  A further one dimensional Bi derived state is observed (Bi3) in the second and third BZ. (b) stack of MDCs cut taken from (a), color ticks mark the peak position of the aforementioned spectral features. The MDCs corresponding to the high symmetry direction $\overline{\Gamma Y}$ ($\overline{X S}$) are red (green) in order to help the readers in tracking the band dispersion in the BZ. The high symmetry directions are more clearly resolved in (c), which displays the MDC at E$_{F}$ along $\overline{X S}$ in the first BZ (red line) and along $\overline{\Gamma Y}$ in the first (black dashed line) and in the second BZ (black solid line). The peak positions, indicated by color ticks, and splittings are different in the different region of the BZ, and they also vary in the different BZs due to the interaction with the substrate. (d) and (e) band dispersion along the chain direction along the $\overline{\Gamma Y}$ and the $\overline{X S}$ high symmetry directions respectively.
(f) calculated band dispersion as in Fig.1(d), but for a wider energy and momentum range. Purple lines are quadratic fits to the Bi3 state. (g) calculated band dispersion along the $\overline{X S}$  high symmetry direction.  
}
  \label{fig:arpes_map}
\end{figure}

The ARPES intensity plot of Fig.~\ref{fig:arpes_map}~(a), shows a constant energy contour at the Fermi energy $E_F$. The Brillouin zones (BZ) of the $p(4\times1)$ reconstruction are outlined by red dashed rectangles. Blue and red arrows point towards the Fermi contours of the Bi1 and Bi2 states, and a green line indicate the bulk Cu Fermi contour. Fig.~\ref{fig:arpes_map}~(b) is a stack of MDCs taken from panel~(a), color ticks mark the peak positions of the Bi derived states and of the Cu \emph{sp} states, in order to help the reader in tracking the dispersion of the bands in the direction orthogonal to the chain direction. Remarkably, the Fermi surface (FS) sheets associated with the spin-split subbands Bi1 and Bi2 are open. This confirms the quasi-1D electronic character of the spin-split states. The Fermi contour of Bi1 suddenly loses intensity outside the 1st BZ. A similar loss of intensity has been observed for states at the Au/Ge(001) $c(8\times2)$ interface, whose dispersion has been proposed to be 1 dimensional \cite{Claessen_PRB_2011}, even though the reduced dimensionality of these states has not been fully confirmed and it is still subject of investigations \cite{Komori_PRB_2011}. Bi2 also becomes weaker, but can still be traced outside the 1st BZ. The contours are clearly not straight, as in an ideal 1D system, but warped. In quasi-1D materials, warping of the Fermi surface is indicative of transverse -- 2D or 3D -- coupling \cite{Grioni_JPCM_2009}. An interaction between the Bi chains, mediated by the Cu substrate, cannot be excluded here, but a closer inspection of the data suggests that the warping may have a different origin.
First of all, the Bi1 and Bi2 contours do not really follow the $(4\times1)$ periodicity of the overlayer in the $\overline{\Gamma\textmd{X}}$ direction, perpendicular to the chains.
Moreover, spectral weight is transferred between the Cu \emph{sp} bulk band and the Bi-derived states in the region of closest approach, around the $\overline{\Gamma\textmd{Y}}$ line. The Bi1 and Bi2 contours appear to be ``squeezed'' towards the center of the BZ, more so than in the following BZs, where the distance from the Cu band is larger.  This effect of transfer of spectral weight between the Brillouin-zones is, as far the Cu bulk band is concerned, similar to the effect in commensurately modulated structures \cite{Voit_Science_2000}, affecting Bi states through the interaction with the substrate bands. Moreover, also other Bi states are affected in this way: a third Bi-induced state (purple line), labelled Bi3, is observed in the second and third BZ. Its contour is also open, but strongly warped. 
The origin of the Bi~3 state can be fully understood by a deeper looking into the calculated band-structure shown in
a wider energy and momentum range in Fig.2(f).  The previously
discussed Rashba-type spin split Bi(1,2) states reach their maximum at 1.3~eV around the $\Gamma$ point. Above these, another pair of Bi $p$-states disperse with negative effective mass, reaching their maximum binding energy at 1.85 eV and crossing at the $\Gamma$ point.
This resembles the situation in the BiCu$_2$ surface alloy on Cu(111) \cite{Bentmann_EPL_2009} (or, similarly,
the BiAg$_2$ alloy \cite{Ast_PRL_2007}), also as far as the different spin-texture is concerned. This
pair of states hybridizes strongly with the Cu states and it is more difficult to follow their dispersion in
the calculation, but an attempt to fit their dispersion results in the purple lines, crossing the Fermi level at $k$-vectors $\sim$ 0.4~\AA$^{-1}$. Hence, we attribute the experimentally observed Bi~3 peak to one of these additional Bi-induced states.


Figure~\ref{fig:arpes_map}~(d) shows the band dispersion parallel to the Bi chains, in the 1st BZ, on a broader energy range than Fig. 1 (b). The interaction with the substrate bulk states is further confirmed by the opening of a hybridization gap in the region enclosed by the black rectangle, near $-1.3$ eV. This is consistent with the calculated band structure of Fig.~\ref{fig:structure_arpes}~(d), which exhibits gaps in the Bi1 and Bi2 dispersion at avoided crossings with the bulk states. Figure~\ref{fig:arpes_map}~(e) and (g) display respectively the measured and calculated dispersion of the one-dimensional states at the border of the first BZ, along the $\overline{\textmd{XS}}$. The measured spin splitting is larger at the zone boundary than at the zone center. Figure~\ref{fig:arpes_map}~(c) displays momentum distribution curves (MDCs) measured at $E=E_{F}$ along the  $\overline{\textmd{XS}}$ high symmetry direction (red line) and along the $\overline{\Gamma\textmd{Y}}$ chain direction, in the first (black dashed line) and second $(4\times1)$ BZ (black continuous line).  The Fermi wave vectors of Bi1 and Bi2, determined by the peak positions (tick marks), are different at the center and at the border of the BZ, and they also vary between the different BZs. The momentum separation of the two subbands is also different. In the 2nd BZ $\Delta$\emph{k} =  0.067 $\pm$ 0.05 \AA$^{-1}$ is $\sim$40\% larger than the corresponding value in the 1st BZ, and also larger than the spin splitting reported at the Bi(111) surface \cite{Koroteev_PRL_2004, Ast_PRL_2007}. The spin splitting is even larger when measured at the border of the BZ, where $\Delta$\emph{k} =  0.075 $\pm$ 0.05 \AA$^{-1}$. It is comparable with the $giant$~spin splitting of the $\mathrm{Bi_{x}Pb_{1-x}Ag_{2}}$ surface alloys \cite{Ast_PRB_2008}. Therefore, the strength of the Rashba effect as extracted from the ARPES data is dependent on the interaction with the substrate. The spin-split states are surface resonance with substrate states whose spectral weight varies in the different Brillouin zones, owing to matrix element effect. As a result, for example, the Bi1 spectral weight fades out near the 1st BZ zone boundary. Therefore, even if the spin-splitting remains in principle constant with $k_{\perp}$, it can appear different in the different Brillouin zone. An additional small variation of the spin splitting between the center zone and the zone boundary arises also from the deviation from a perfectly one-dimensional state.  


\begin{figure}[!t]
  \centering
 \includegraphics[width = 1.0\textwidth]{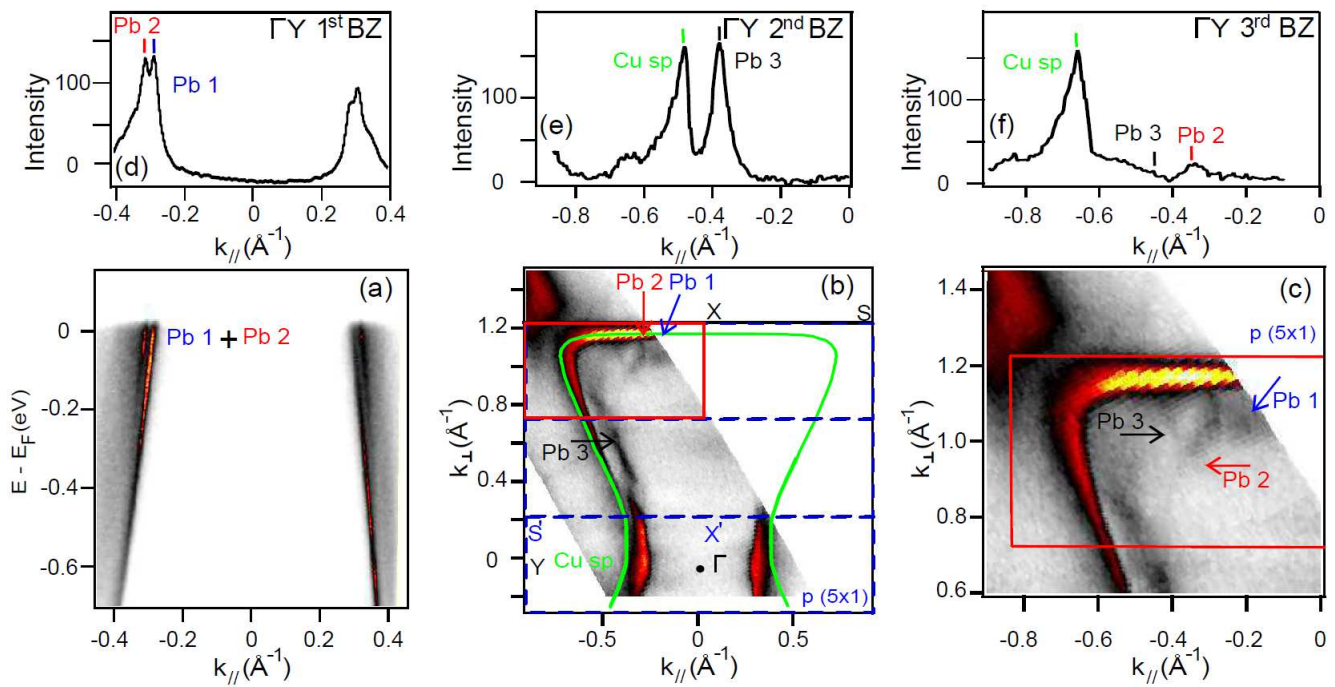}
  \caption[didascalia]{(a) Experimental band dispersion of the spin-split surface states of the Pb/Cu(110) $p(5 \times 1)$ interface along the $\overline{\Gamma Y}$ high symmetry direction. (b) ARPES Fermi surface contour. Blue dashed rectangles indicate the $p(5 \times 1)$ BZs; color arrows indicate the contours of the various states. (c) close up of the shaded region in (b). (d-f) MDCs at $E_{F}$ along $\overline{\Gamma Y}$ in three different BZs. Color ticks indicate the Fermi wave-vector of the different bands.}
  \label{fig:PbCu}
\end{figure}


\section{Electronic properties of Pb/Cu(110) $\mathbf{p(5\times1)}$ and of the mixed surface alloy}

The contribution of the atomic SOC to the large Rashba splitting discussed in the previous section can be assessed by substituting bismuth by lead. Pb is an ideal choice because: i) it has a large atomic SOC; and ii) the atomic radii of Bi and Pb are quite close ($3.1~\AA$ for the Bi - Bi distance \cite{Lottermoser_SS_1996} and $2.9~\AA$ for the Pb - Pb distance \cite{Huang_NJP_2011}), resulting in a very similar surface reconstruction. The phase diagram of the Pb-Cu(110) interface is complex in the sub-monolayer (ML) coverage range \cite{marra_PRL_1982, Brennan_PRB_1986, Beau_PRB_1991, Nagl_PRB_1995}. Similarly to the case of Bi, a $c(2 \times 2)$ superstructure is formed at $\sim$~0.5~ML coverage. For larger coverages, several 1D $p(n \times 1)$ phases are found, with different values of the interchain spacing. They arise from the substitution of every n-th row of Cu with Pb atoms, similarly to the case of the Bi/Cu(110) $p(4 \times 1)$ interface shown schematically in Fig.~\ref{fig:structure_arpes}~(b). We report here ARPES results for the $p(5 \times 1)$ phase.

Figure~\ref{fig:PbCu}~(a) displays the band dispersion along the $\overline{\Gamma\textmd{Y}}$ direction, parallel to the Pb chains. Two intense and sharp states are resolved, labelled Pb1 and Pb2 in analogy with the Bi case. They cross the Fermi level at $k_{F1} = \pm 0.31$ \AA$^{-1}$ and $k_{F2} =  \pm  0.28$ \AA$^{-1}$, and the splitting measured in the 1st BZ is $\Delta k = 2 k_{0} = 0.029$ \AA$^{-1}$. The larger $k_{F}$ values with respect to the Bi/Cu(110) $p(4 \times 1)$ case reflect the different band filling for the 4 (5) valence electrons of Pb (Bi), in a rigid band scenario. 
The spin-splitting is reduced by a factor $1.7$, to be compared with the ratio of $1.37$ of the atomic SOC in Bi and Pb. A similar reduction was observed for the $\mathrm{BiAg_{2}}$ and $\mathrm{PbAg_{2}}$ surface alloys \cite{Ast_PRB_2008, Pacile_PRB_2006}. It shows that the atomic SOC is certainly an important factor, but certainly not the only element determining the strength of the 1D Rashba effect, which is the result of a complex interplay between atomic and structural parameters, similar to the case of the 2D surface alloys.

Figure~\ref{fig:PbCu}~(b) shows an ARPES Fermi contour, to be compared with Fig.2 (a). Blue dashed rectangles define the $p(5 \times 1)$ BZs. The closed contour (green) outlines the FS of the bulk Cu \emph{sp} conduction band. The Pb1 and Pb2 spin-split states (indicated by color arrows) and a third Pb derived band (Pb3), exhibit open contours. Similarly to the Bi case, the contours are warped, and the two spin-split states rapidly lose intensity beyond the 1st BZ. Again, this can be seen as the result of the interaction of the 1D interface states with the substrate. The concavity of the third Pb derived band, Pb3, is opposite to that of the spin-split states, in agreement with the Bi case.


\begin{figure}[!t]
  \centering
 \includegraphics[width = 0.9\textwidth]{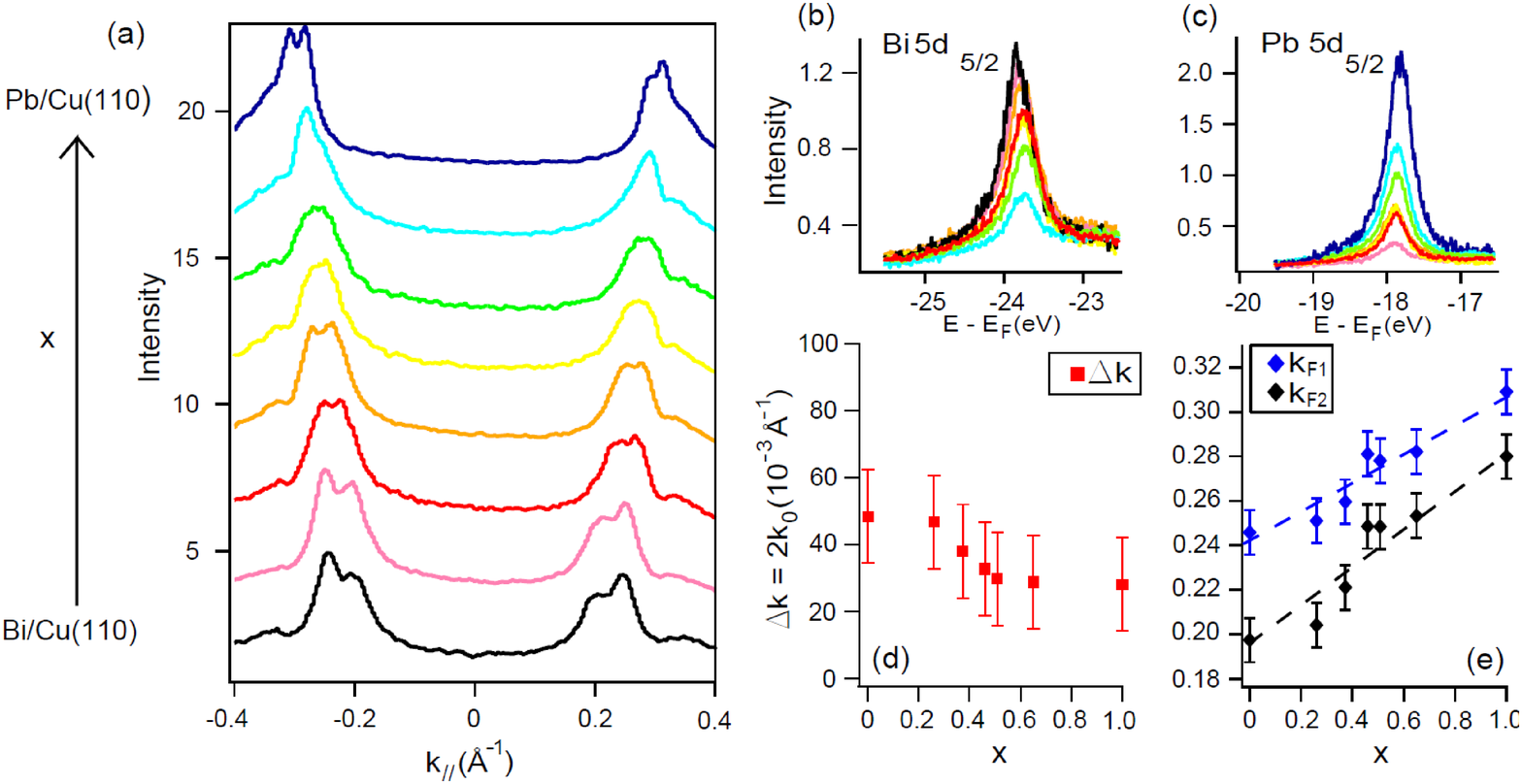}
  \caption[didascalia]{(a) MDCs measured at $E=E_F$ along $\overline{\Gamma Y}$ in the 1st BZ for various $\mathrm{Bi_{1-x}Pb_{x}/Cu(110)}$ mixed alloys. The relative concentration of Pb and Bi was estimated from the $5d_{5/2}$ core levels, as shown in (b) and (c) respectively for Bi and Pb. The $k_{f}$ values clearly increase with \emph{x}, as shown in detail in panel (e), where the peak position is extracted from a multi-Lorentzian fit. (d) shows the corresponding reduction of the spin splitting.}
  \label{fig:mixed}
\end{figure}


We investigated in detail the 1D Pb-derived states in the 3rd $(5 \times 1)$ BZ, where the interaction with the substrate is weakest. In Fig. 3 (c) the contour of Pb2 is more clearly resolved. The 2D warping at the Fermi level is $\pm0.05~$\AA$^{-1}$, with a $15\%$ modulation of the $k_{F}$ value. These values are larger than the corresponding values for the Bi case ($\pm0.03~$ \AA$^{-1}$ and $10\%$). Since the interchain distance in the $p(5 \times 1)$ phase is $25\%$ larger ($1.27~nm$ vs. $1.02~ nm$), we conclude again that the overlayer-substrate interaction, rather than the transverse interchain coupling, is the main cause for the warping of the 1D FS. Figure~\ref{fig:PbCu} (d-f) display MDCs measured at $E=E_{F}$ along the $\overline{\Gamma Y}$ direction in the three different BZs. Color ticks mark the Fermi wave vectors of the various bands. The position of Pb1 is hardly detectable in the higher order BZs, where Pb3 acquires larger spectral weight. In the 2nd BZ (e) in particular it is difficult to resolve Pb3 from Pb2, since the two almost merge as it is visible also in panel (b), but the two are better visualized in the 3rd BZ in panels (c) and (f).

The possibility to manipulate the electronic properties of the $p(n \times 1)$ phase was explored in the mixed $\mathrm{Bi_{1-x}Pb_{x}/Cu(110)}$ surface alloy. The similar atomic radii and the comparable surface free energies result in similar surface reconstructions, and thus enabled us to study the whole range of stoichiometries between $x=0$ and $x=1$. Figure~\ref{fig:mixed}~(a) shows  momentum distribution curves (MDCs) measured at $E_{F}$ along $\overline{\Gamma Y}$ in the 1st BZ. The spectra are vertically offset for clarity, with the pure Bi/Cu(110) (Pb/Cu(110)) at the bottom (top). The MDCs clearly indicate that the increasing of \emph{x} induces a continuous rigid shift in binding energy of the surface states and a reduction of the spin splitting. Figure~\ref{fig:mixed}~(b) and (c) shows the Bi and Pb $5d_{5/2}$ core level spectra used to determine the relative concentration of Bi and Pb. 
To be more quantitative, we fitted each MDC by two Lorentzian doublets (one on each side of $k=0$), describing the spin branches, and a background to account for contribution of the Cu \emph{sp} band. The peak positions of the Lorentzians, corresponding to $k_{F1}$ (outer branch) and $k_{F2}$ (inner branch), are shown in Figure~\ref{fig:mixed}~(e) as a function of \emph{x} and the dashed lines are guides to the eye. The difference between each pair of $k_{F}$ values defines the spin-splitting, shown in panel (d). A clear reduction of $\Delta k = 2 k_{0}$ is observed with increasing Pb content.


\section{IV. Conclusion}

In summary, we have found and characterized by ARPES one dimensional spin-split surface states at the Bi/Cu(110) $p(4\times1)$ and Pb/Cu(110) $p(5\times1)$ interfaces. The experimental findings are supported by DFT calculations including spin-orbit coupling effects. The quasi-1D electronic character of these states reflects the structurally 1D character of parallel chains formed by Bi atoms in substitutional sites. The large buckling of the Bi atoms with respect to the substrate is a common element with the 2D alloys grown on the (111)-terminated surface of Ag and Cu \cite{Gierz_PRB_2010}. The \emph{k} splitting is larger than the values reported for the surface states of Au(111) \cite{Lashell_PRL_1996} and Bi(111) \cite{Koroteev_PRL_2004}, and it is comparable with the giant spin splitting observed in the Ag(111) surface alloys \cite{Ast_PRL_2007, Ast_PRB_2008, moreschini_PRB_2009, Bentmann_EPL_2009}. Furthermore, we observe a difference in the spin splitting going from the first to higher-order Brillouin zones. This is interpreted as a consequence of the hybridization between the 1D interface states and the bulk continuum. It suggests that the measured large spin splitting strongly depends on the properties of the surface state wavefunctions and their interaction with the substrate.

We have addressed the role of the atomic SOC in defining the magnitude of the spin splitting by substituting Bi with Pb. We investigated the Pb/Cu(110) $p(5\times1)$ interface, and we report a reduction of the spin splitting equal to $\sim 1.7$, which is larger than the ratio between the atomic SOC ($\sim 1.37$). This indicates that also in these one dimensional surface alloys the surface structure, namely the buckling of the heavy atoms in substitutional site, plays an important role in determining the value of the spin splitting. 

A complementary point of view would consist in measuring the evolution of the spin splitting for the different Pb $p(n \times 1)$ interfaces, in order to quantify the role played by the surface structure. The Pb/Cu(110) system might play an important role in isolating the structural influence (namely the buckling $\Delta z$) on the spin splitting, without need to change the adsorbed atom but simply by modifying \emph{n}. A systematic combined high resolution ARPES and structural (XRD, or XPD or IV-LEED) investigation would be desirable. 

\section*{References}

\providecommand{\newblock}{}

\end{document}